\documentclass{article}

\pdfoutput=1

\usepackage{arxiv}

\usepackage[utf8]{inputenc} % allow utf-8 input
\usepackage[T1]{fontenc}    % use 8-bit T1 fonts
\usepackage{hyperref}       % hyperlinks
\usepackage{url}            % simple URL typesetting
\usepackage{booktabs}       % professional-quality tables
\usepackage{amsfonts}       % blackboard math symbols
\usepackage{nicefrac}       % compact symbols for 1/2, etc.
\usepackage{microtype}      % microtypography
\usepackage{cleveref}       % smart cross-referencing
\usepackage{lipsum}         % Can be removed after putting your text content
\usepackage{graphicx}
\usepackage{doi}

\usepackage{hanging} % for manual References section

\usepackage[flushleft]{threeparttable}
\usepackage{graphicx}
\graphicspath{ {figures/} }
\usepackage{longtable}

\usepackage{titlesec}
\titlespacing{\section}{0pt}{\parskip}{-\parskip}
\titlespacing{\subsection}{0pt}{\parskip}{-\parskip}
\titlespacing{\subsubsection}{0pt}{\parskip}{-\parskip}

\title{Truth in Text: A Meta-Analysis of ML-Based Cyber Information Influence Detection Approaches}

\date{}

\author{Jason M. Pittman\\
	University of Maryland Global Campus\\
	\texttt{https://orcid.org/0000-0002-5198-8157} \\
}

\hypersetup{
pdftitle={Truth in Text: A Meta-Analysis of ML-Based Cyber Information Influence Detection Approaches},
pdfsubject={},
pdfauthor={Jason M. Pittman},
pdfkeywords={cyber information influence, misinformation, disinformation, detection, machine learning, artificial intelligence},
}

\begin{document}
\maketitle

\begin{abstract}

Cyber information influence, or disinformation in general terms, is widely regarded as one of the biggest threats to social progress and government stability. From US presidential elections to European Union referendums and down to regional news reporting of wildfires, lies and post-truths have normalized radical decision-making. Accordingly, there has been an explosion in research seeking to detect disinformation in online media. The frontier of disinformation detection research is leveraging a variety of ML techniques such as traditional ML algorithms like Support Vector Machines, Random Forest, and Naïve Bayes. Other research has applied deep learning models including Convolutional Neural Networks, Long Short-Term Memory networks, and transformer-based architectures. Despite the overall success of such techniques, the literature demonstrates inconsistencies when viewed holistically which limits our understanding of the true effectiveness. Accordingly, this work employed a two-stage meta-analysis to (a) demonstrate an overall meta statistic for ML model effectiveness in detecting disinformation and (b) investigate the same by subgroups of ML model types. The study found the majority of the 81 ML detection techniques sampled have greater than an 80\% accuracy with a Mean sample effectiveness of 79.18\% accuracy. Meanwhile, subgroups demonstrated no statistically significant difference between-approaches but revealed high within-group variance. Based on the results, this work recommends future work in replication and development of detection methods operating at the ML model level.

\end{abstract}


\section{Introduction}
\label{sec:introduction}
%% short paragraph introducing cyber information influence in the context of cybersecurity
Cyber information influence refers to the strategic use of digital platforms to shape or control behavior. Such influence can be leveraged for legitimate purposes, such as public awareness campaigns, but it is also a critical tool in cyber-enabled disinformation efforts. In the latter context especially, cyber information influence is a growing cybersecurity threat. Adversaries, including state actors, cybercriminals, and hacktivists, exploit social media algorithms, artificial intelligence, and bot networks to manipulate narratives, destabilize societies, and undermine trust in institutions. Furthermore, cyber information influence campaigns often coincide with cyberattacks, such as data breaches, ransomware attacks, and infrastructure disruptions, to create chaos, reduce resilience, and erode public confidence in digital systems.

Indeed, the rapid proliferation of cyber information influence (i.e., misinformation, disinformation, fake news) across digital platforms has emerged as a significant societal challenge. The effects are wide-ranging with implications to public opinion, institutional trust, and social cohesion [7, 56, 36]. These effects emerge because of false information spreading quickly due to the viral nature of online content and the psychological tendencies of users to engage with emotionally charged material. Such activity further exacerbates the challenge of information containment [30]. In response, researchers have developed a variety of machine learning (ML) techniques to detect the various types of cyber information influence. Put simply, ML-based detection techniques provide key indicators of false or misleading information [11, 37, 15].

Despite significant advancements however, the field of cyber information influence detection remains fragmented, with conflicting findings on the effectiveness of different ML approaches. Some studies suggest that deep learning models, such as Convolutional Neural Networks (CNNs) and Long Short-Term Memory (LSTM) networks, outperform traditional methods due to the ability to capture complex semantic relationships in text [34, 20]. Other research highlights the continued relevance of traditional ML models, such as Support Vector Machines (SVMs) and Random Forest classifiers. Such traditional models may provide comparable or superior results with lower computational costs [10, 30]. Additionally, inconsistencies in datasets, evaluation methodologies, and feature engineering techniques contribute to the lack of consensus regarding a generalized effective misinformation detection strategy [33, 43, 49].

To address these challenges, this study conducted a three-phased meta-analysis of existing ML-based techniques for cyber information influence detection. By systematically synthesizing findings from peer-reviewed literature, this research aimed to provide a precise estimate of detection effectiveness, identify trends across different ML methodologies, and uncover factors that influence performance outcomes [42, 44, 55]. A key goal was to standardize performance reporting across studies to facilitate more consistent and reliable comparisons. Ultimately, doing so contributes to a deeper understanding of sentiment analysis effectiveness in cyber information influence- specifically, disinformation- detection.

The remainder of this paper is structured as follows. The body of related work relevant to situating this study in the existing research is presented in the Related Work. Then, the Methodology section outlines the methodology used to conduct the meta-analysis, detailing the search strategy, inclusion criteria, and data extraction process. This is followed by the Results section which presents and discusses the statistical analysis. Finally, the Conclusions and Future Work section interprets the implications of the findings, and concludes the paper with a summary of key contributions and recommendations for future research.

\section{Related Work}
\label{sec:related-work}
The proliferation of cyber information influence across digital platforms poses significant challenges in various domains, including public health, politics, and finance [37, 55]. Further, the notion that online discourse or shared information might be of dubious value distorts social norms. This is because cyber information influence, while once exclusive to nation state soft power doctrine, has filtered down to a variety of technology augmented social strata. Certainly, large campaigns require state level resources and coordination. However, individual social media users as well as small ideologically aligned groups can exert cyber information influence across a plethora of knowledge domains. 

\subsection{Misinformation, Disinformation, and Fake News}
Cyber information influence is normally understood to consist of three categories. While misinformation refers to the unintentional spread of false or misleading information, disinformation is the deliberate dissemination of deceptive content with the intent to mislead audiences. Fake news, then, is disinformation masquerading as legitimate news. The increasing accessibility of social media and the ease of content generation have exacerbated the rapid spread of disinformation in specific, overwhelming traditional fact-checking efforts that rely on manual verification by human experts [40]. As a result, researchers have explored automated solutions to enhance the detection and mitigation of cyber information influence [15, 42, 43].

\subsection{Evolution of Detection Techniques}
Broadly speaking, there are four groups of cyber information influence detection techniques (Table \ref{table:1}). Each represents a step in the overarching chronology of technological advancement. 

\begin{table*}[!htbp]
  \centering
  \begin{threeparttable} \caption{Summary of Cyber Information Influence Detection Approaches}
      \begin{tabular}{lccc}
          \toprule
          {\textbf{Approach}} & \textbf{Strengths} & \textbf{Weaknesses} & \textbf{Best Use Case} \\
          \hline
          \hline
          Rule-based & Simple & Rigidity & Structured \\
          & Interpretable & High false-positive rate & Well-defined \\ \hline 
          Traditional ML & Scalable & Context limited & News \\
          & Data-driven & Requires feature engineering & Structured text \\ 
          & Accurate & & \\ \hline 
          Deep learning & Context-aware & Opaque & Social media \\
          & Feature automation  & Compute expensive & Large datasets \\ \hline 
          Hybrid & Context-rich detection & Resource intense & Multi-platform or media \\
          & Complex integration & & \\
          \bottomrule
      \end{tabular}
      \label{table:1}

  \end{threeparttable}
\end{table*}

\subsubsection{Rule-Based}
Early cyber information influence detection efforts primarily employed rule-based approaches [48], which relied on handcrafted linguistic features, keyword matching, and syntactic patterns to flag content [46, 65]. Such approaches utilized predefined dictionaries, syntactic patterns, and stylistic features to target characteristics such as sensational language, clickbait phrases, and linguistic anomalies [28, 44]. However, rule-based methods suffer from three significant limitations. Foremost, such methods are rigid. In other words, rules-based approaches lack adaptability to evolving misinformation tactics [30, 36]. Consequently, rules-based techniques tend to exhibit high false-positive rates. These methods often flag legitimate content incorrectly, especially when evaluating data even moderately outside the initial configuration or rule parameters. Thus, a third limitation can be contextualized as domain dependency. That is, as with any rule-drive system, the rules must be constantly maintained if the system is to retain effectiveness in its task. As a result, rule-based approaches were gradually replaced by more flexible ML techniques that could learn patterns directly from data [34, 45, 47].

\subsubsection{Traditional ML}
To overcome the rigidity of rule-based systems, researchers adopted ML approaches. ML introduced statistical methods to automatically learn patterns from data. Early attempts employed ML models such as SVM, DT, and RF leveraged handcrafted features such as n-grams, readability scores, and sentiment polarity to differentiate between legitimate and misleading content [12]. As a general method, ML addresses the rigidity found in rules-based approaches thereby obviating domain dependency to various degrees of success. ML also scales better without the need for hands-on maintenance requirements. Yet, for all the advantages compared to rule-based methods, ML brought along a new set of limitations. Effectiveness and performance depend on careful feature engineering by human operators. Therefore, appropriate domain knowledge is necessary to engineer a robust and reliable implementation. Moreover, a nontrivial volume of problem-space specific data is required to properly train and tune ML detection techniques. This renders the ML model tightly coupled to both the selected features and problem definitions. Such is not conceptually different from the rules-based limitations insofar as the result is a nontrivial false positive rate. Further, ML has contextual understanding limitations because models cannot capture semantic relationships within text. These shortcomings led to the adoption of deep learning models, which offered automatic feature learning capabilities [44, 45].

\subsubsection{Deep Learning}
Enter deep learning models. These include Convolutional Neural Networks (CNNs), Long Short-Term Memory (LSTM) networks, and Transformer-based architectures [32] (e.g., BERT, RoBERTa). These more modern approaches mark a significant advancement in cyber information influence detection [41]. To that end, deep learning models have demonstrated superior performance largely due to contextual embeddings and attention mechanisms capturing the subtle patterns that traditional ML approaches fail to handle. Deep learning also eliminates the need for manual feature engineering. However, challenges persist, such as requiring large, labeled datasets to achieve high performance [65]. Furthermore, deep learning has high computational complexity and high resource requirements. Both limit deployment feasibility. Meanwhile, deep learning models are black box in nature making it difficult to understand model decisions [35, 63, 64]. One can reasonably deduce then that despite the strengths of deep learning, the method alone is not sufficient for robust misinformation detection. Hence, the exploration of hybrid approaches incorporating additional contextual signals [40].

\subsubsection{Hybrid}
To address the limitations of content-based approaches, recent research has shifted towards hybrid models, which combine textual analysis with auxiliary features such as social network structures, temporal patterns, and user credibility scores [41, 49]. The motivation is to assess the reliability of content sources in addition to tracing user engagement patterns through interaction graphs. Overall, hybrid models have demonstrated improved robustness by leveraging multimodal data. Indeed, hybrid approaches have improved detection accuracy compared to traditional and deep learning techniques. However, hybrid methods introduce additional complexity and require significant computational resources [12, 20, 33].

\subsection{Need for Meta-Analysis}
Despite significant advancements in cyber information influence detection techniques over time, one can observe three persistent challenges across all approaches. The literature demonstrates variability in datasets, feature selection, and evaluation metrics across studies. The variability limits the ability for one to determine the most appropriate technique for a given context. High volumes of data can be a challenge for real-time detection schemes as well. Moreover, even when a technique exists capable of such processing, the computational demands scale in lockstep which makes experimentation, let alone production implementations, costly.  Like the constraints associated with rigidity or tight coupling to training datasets observed in earlier techniques, deep learning models often perform well on benchmark datasets but struggle in applied, practical scenarios. Further, existing literature focuses on qualitative assessments of detection technique performance with broad variance in quantitative result reporting. Often, the literature presents conflicting results due to variations in study design and datasets used. To that end, a meta-analysis is crucial to quantitatively assess the effectiveness of different AI-based misinformation detection techniques, identify consistent trends, and provide evidence-based recommendations for future research [20, 28].

\section{Methodology}
The purpose of this research was to conduct a comprehensive meta-analysis of existing ML-based techniques for the detection of cyber information influence. This included all commonly understood forms: misinformation, disinformation, and fake news. The work aimed to synthesize quantitative performance findings across various ML approaches. Goals included demonstrating a unified measure, identifying key trends and methodological differences, and providing precise assessments of detection effectiveness. Furthermore, by aggregating results from peer-reviewed literature, this analysis sought to address perceived inconsistencies in reported performance between studies, improve the precision of current estimates, and inform future research directions.

All the above were encapsulated in two research questions. The first was, \textit{what is the overall effectiveness, measured in model accuracy, of ML-based cyber information influence detection}? The second was, \textit{what is the effectiveness, measured in model accuracy, of each ML approach and to what extent are there significant differences between approaches}?

To affect such as purpose, and to investigate potential answers to these questions, the study employed a meta-analysis methodology. More specifically, a three-stage approach assessed nine ML-based detection literature reviews yielding 40 unique studies as the secondary data corpus. The corpus contributed 153 ML accuracy values across 81 model types and 37 unique datasets. The first stage involved an overall assessment of ML-based detection techniques. This imparted an overarching perspective of ML-based technique performance. Then, because of the variety of ML models used throughout the collected literature, combined with the lack of uniformity in reported metrics and diversity of datasets, the second stage analyzed studies by specific ML approaches (e.g., between-approaches). Doing so provided a more granular analysis based on factors per subgroups. Furthermore, such localization worked to uncover the most effective techniques and reveal how variations in the corpus methodologies impacted detection performance. Finally, this work concluded with statistical testing of potential significant differences between ML model type subgroups along with comparative inferences. 

\subsection{Overview of the Meta-Analysis Approach}
Meta-analysis is an established quantitative research method that aggregates findings from multiple independent studies to provide a more reliable and generalizable conclusion [42]. The method was well-suited for this study due to several study characteristics. First, resolving inconsistencies, filling quantitative gaps, and deconflicting results in the existing literature is paramount. Doing so tracks with previous studies on misinformation detection having reported varying performance outcomes [34, 44]. To that end, meta-analysis enabled a systematic synthesis of the corpus results while imparting consistency to trends and smoothing out gaps and discrepancies. Furthermore, enhancing statistical power and precision is a central feature of meta-analysis. 

The selected cyber information influence detection literature exhibited limited sample sizes and thus demonstrated low statistical power and non-generalizable conclusions [49]. By aggregating data across studies, meta-analysis increased precision and provided more robust effect size estimates. Similarly, meta-analysis is appropriate for standardizing ML performance metrics. Given the observable challenge in misinformation detection research related to the use of diverse evaluation metrics (e.g., accuracy, precision, recall, F1-score, AUC-ROC), meta-analysis was uniquely positioned to develop a meta-statistic to ease comparison [41]. In this way, meta-analysis facilitated the harmonization of metrics and established consistency in potential interpretations.

Moreover, generalizing across different contexts is also a strength of meta-analysis. Cyber information influence detection techniques are applied to various data sources in the literature including social media, news articles, and user comments, with each presenting unique challenge (Puska et al., 2024). A meta-analysis helped assess the generalizability of techniques across different domains. Finally, meta-analysis facilitated the identification of knowledge gaps and future research directions. Through systematic review and effect size estimation, meta-analysis uncovers underexplored areas and informs future research priorities [41].

\subsection{Search Strategy and Data Collection}
A comprehensive and systematic search strategy was employed to identify relevant studies for this meta-analysis. The search process was designed to ensure the inclusion relevant research while minimizing selection bias. To achieve this, an extensive search was conducted across three major academic databases: Google Scholar, IEEE Xplore, and the ACM Digital Library. These databases were selected for their broad coverage of artificial intelligence, machine learning, and computational cyber information influence research. The search terms were formulated to encompass three primary thematic areas: cyber information influence concepts, machine learning methodologies, and literature type. Boolean search queries were constructed using combinations of terms such as \textit{misinformation}, \textit{disinformation}, \textit{fake news}, \textit{machine learning}, \textit{deep learning}, \textit{transformers}. Additionally, both \textit{review}, and \textit{survey}, were operationalized as part of the search to ensure that only comparative or meta-analytic studies were retrieved.

To maintain relevance to current research trends, studies were restricted to those published between 2016 and the present. Studies contributing accuracy values to the meta-analysis were not so constrained. Non-peer-reviewed sources such as blog posts and opinion pieces were excluded. Similarly, studies without a strong technical focus (e.g., those addressing policy or psychological aspects of misinformation without applying machine learning methodologies) were omitted.

The study selection process followed the Preferred Reporting Items for Systematic Reviews and Meta-Analyses (PRISMA) guidelines, ensuring transparency and reproducibility. The final dataset included 40 unique studies, extracted from nine systematic reviews, yielding 153 distinct ML accuracy values across 81 model types and 37 datasets.

\subsection{Data Analysis}
Operationalizing the above allowed this work to address the research problem wherein there is lack of generalized misinformation detection effectiveness in the literature. This was accomplished in three phases. The first phase provided an overarching analysis across all ML-based disinformation detection techniques. The second and third phases probed deeper into specific groups of techniques and examined potential statistically significant differences.

Data analyses were conducting using instrumentation developed in Python. Discrete programs completed each meta-analysis phase separately using the comma-separated value data extracted from the sample literature. Results were either written to plaintext files (tables) or used to generate plots (figures). Standard packages were employed: pandas, matplotlib, numpy, and scipy.stats.

The first phase treated extracted accuracy values ($n$ = 153) as individual effect sizes. Given the sample studies did not uniformly report sample sizes or within-study variance, an unweighted approach was used. Specifically, each accuracy value was taken as equally informative. The pooled estimate was computed as the Mean of these values. The study also calculated a 95\% confidence interval via the standard error of the mean. To visualize the data, a forest-style plot listed the accuracy measures in ascending order, alongside a vertical line denoting the pooled mean. Although this method does not weight entries by sample size, Phase 1 provided an aggregate snapshot of general effectiveness across the sample of detection techniques as well as an answer to the first research question.

Because the literature encompassed a wide range of machine learning algorithms (e.g., SVM, CNN, BERT, and so on), Phase 2 disaggregated the meta-analysis by approach category. The phase leveraged a string-matching procedure to classify each detection technique (for instance, roBERTa or BioBERT mapped under BERT). This yielded well-defined subgroups such as SVM, LR, NB, LSTM, and so forth. Each subgroup’s effect sizes were then aggregated separately, again under an unweighted scheme. For each category, the study again computed the Mean accuracy, Standard Deviation, and a 95\% confidence interval. A robust summarization table accompanied this phase and illustrated the Mean ± CI for visual comparison of approach-level performance. In doing so, Phase 2 isolated performance tendencies specific to each method family and highlighted potential leaders or underperformers in the misinformation detection landscape. 

Finally, Phase 3 employed a one-way analysis of variance (ANOVA) treating the approach category as the independent variable and accuracy as the dependent variable. Doing so enabled a formal test of whether at least one group differed significantly from the others. We executed the ANOVA under an ordinary least squares (OLS) framework, extracting degrees of freedom, sum of squares, and mean squares for complete transparency. If a significant omnibus result emerged, the study would leverage Tukey’s HSD to determine which specific algorithm pairs exhibited statistically meaningful gaps in performance. This phase thus moved beyond overarching effectiveness and addressed the study’s second research question- namely, to what extent different ML approaches truly diverge in detection effectiveness.

\section{Results and Discussion}
\begin{figure}[h]
  \centering
  \caption{The distribution of studies per analyzed review}
  \label{fig:1}
  \includegraphics[scale=0.6]{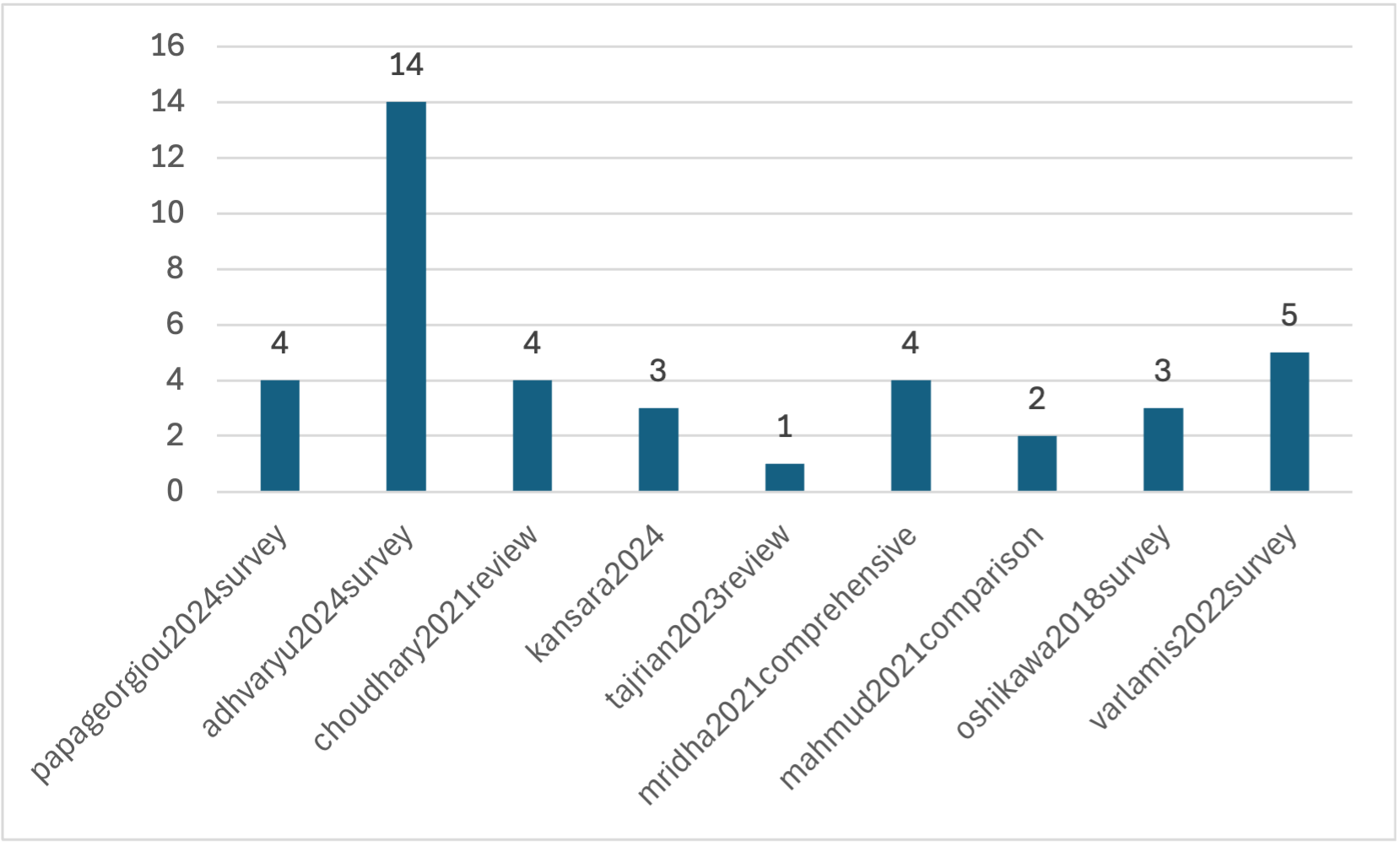}
\end{figure}

The results are presented according to the three phases of the meta-analysis in the following three sections. A total of 40 studies were analyzed to evaluate the effectiveness (e.g. accuracy) of machine learning-based disinformation detection techniques based on nine source literature reviews (Figure \ref{fig:1}). Accuracy metrics for ML-based detection techniques were extracted and cataloged in a simple comma-separated value file along with the study bibtex citekey, ML model type, and dataset, and accuracy in decimal format. Data were then cleaned to ensure consistent representation in terms of spelling, column placement, accuracy score decimal places and rounding, and de-duplication. These data were then analyzed using the developed Python instrumentation.

Descriptively, the sample studies were evenly distributed across the nine foundational review papers with two exceptions. One [1] review contained a significant percentage of model accuracy results (35\%). On the opposite, another [49] review yielded a single result. The remaining six foundational reviews provided between 5\% and 12\% of the accuracy value sample. 

\begin{figure}[h]
  \centering
  \caption{The distribution of ML models in the studies analyzed}
  \label{fig:2}
  \includegraphics[scale=0.5]{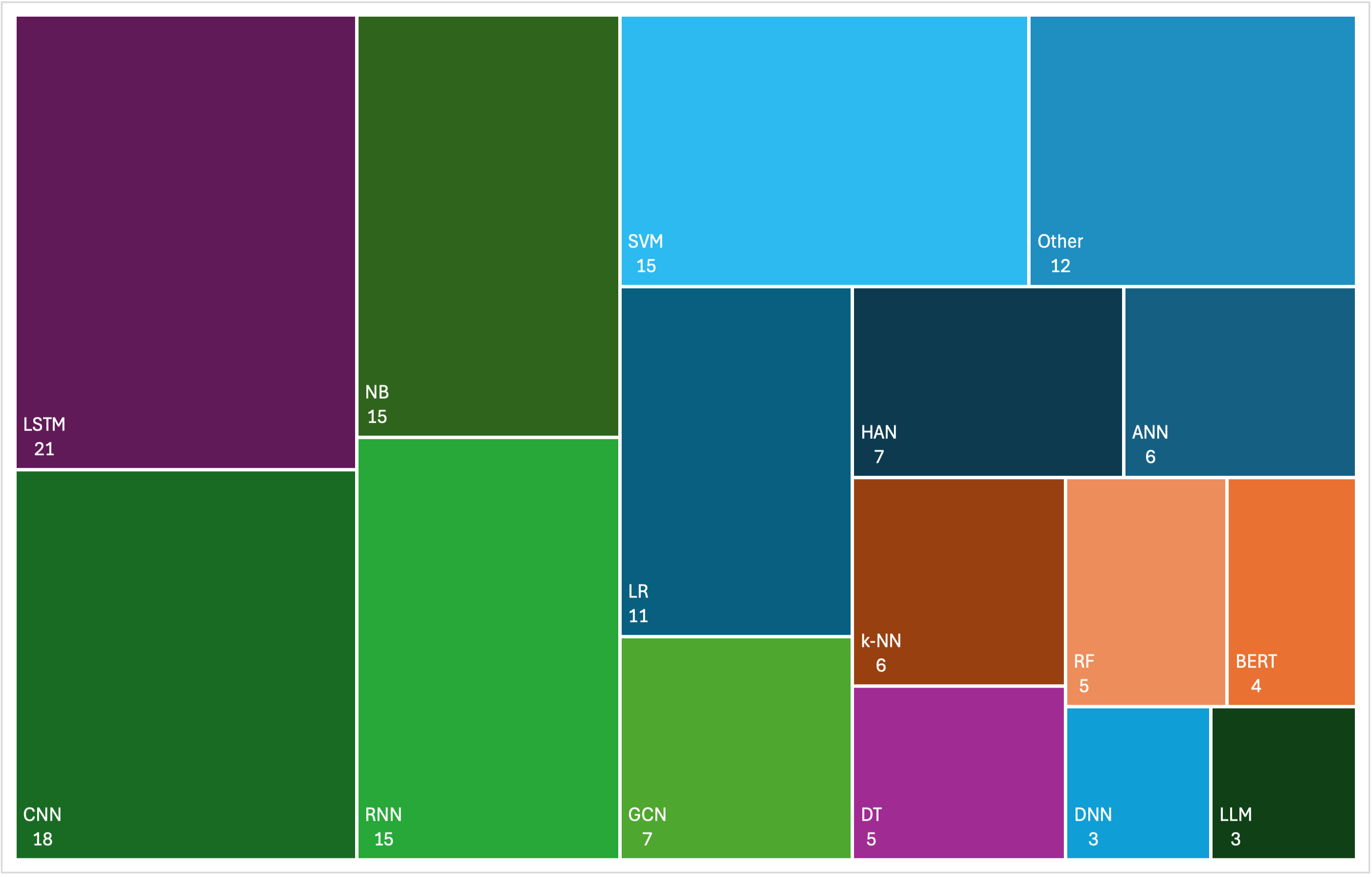}
\end{figure}

Relatedly, a core set of ML model types appeared throughout the sample (Figure \ref{fig:2}) with some minor exceptions representing frontier techniques or novel detection attempts. Notably, there was a rather even observed mixture between traditional ML approaches and deep learning-based techniques. Large Language Model (LLM) methods represented the lowest frequency while CNN based techniques along with LSTM methods yielded the highest frequencies. Interspersed throughout the remainder of the accuracy value sample were traditional models such as NB and SVM along with deep learning models such as RNN.

\subsection{Phase 1: Overall Meta-Analysis}

\begin{table*}[!htbp]
  \centering
  \begin{threeparttable} \caption{Overall (Unweighted) Meta-Analysis Results}
      \begin{tabular}{lc}
          \toprule
          {\textbf{Measure}} & \textbf{Value} \\
          \hline
          \hline
          Number of Accuracy Metrics ($n$)	& 153 \\
          Mean Accuracy &	78.19\% \\
          95\% CI	& [75.381\%, 80.999\%] \\
          Std. Dev &	17.668\% \\
          Median Accuracy &	82.36\% \\
          Interquartile Range (Q1-Q3) &	68.302\% - 91.61\% \\
          Min/Max Accuracy &	23.3\% / 99.9\% \\
          \bottomrule
      \end{tabular}
      \label{table:2}
  \end{threeparttable}
\end{table*}

The aim of the first phase of the meta-analysis was to investigate a generalized effectiveness picture for the state of ML-based detection techniques. Doing so worked towards answering the first research question which inquired about the overall effectiveness of these techniques. To that end, this phase calculated a set of descriptive statistics (Table \ref{table:2}) based on the sample of accuracy performance metrics ($n$ = 153). The Mean and Median accuracy results show an overarching effectiveness. However, adjacent results such as Standard Deviation, Interquartile Range, as well as Minimum and Maximum values reveal a wide dispersion of accuracy metrics in the sample. Yet, the Confidence Interval indicates a 95\% potential for the population Mean accuracy to exist close to the Mean and Median range.

Two additional descriptions of ML model accuracy performance provide a richer analysis. First, a distribution of accuracy values (Table \ref{table:3}) shows occurrences of absolute and relative accuracy values. The description indicates most (i.e., $>$ 5\%) misinformation detection techniques have a score at or above 71\% to 100\% accuracy. Moreover, only 9.15\% of the sample values were equal to or below 50\% accuracy. Indeed, the highest frequency of values exhibited over 90\% accuracy ($n$ = 45).

\begin{table*}[!htbp]
  \centering
  \begin{threeparttable} \caption{Frequency distribution of accuracy values}
      \begin{tabular}{lcc}
          \toprule
          {\textbf{Accuracy}} & \textbf{Absolute}  & \textbf{Relative}\\
          & \textbf{Frequency} & \textbf{Frequency} \\
          \hline
          \hline
          0 – 10\%  &	0 & 0 \\
          11 – 20\% &	0	& 0 \\
          21 – 30\% & 4 &	2.61\% \\
          31 – 40\% & 2 &	1.31\% \\
          41 – 50\%	& 8	& 5.23\% \\
          51 – 60\%	& 15  &	9.8\% \\
          61 – 70\% & 12  & 7.84\% \\
          71 – 80\% & 25	& 16.34\% \\
          81 – 90\% & 42  & 27.45\% \\
          91 – 100\%  & 45  & 29.41\% \\
          \bottomrule
      \end{tabular}
      \label{table:3}
  \end{threeparttable}
\end{table*}

Then, lastly, a full descriptive breakdown in Appendix B demonstrates a per study (indicated by bibtex citekey), per ML model, accuracy relative to the sample Mean. This breakdown, in relationship to Appendix A, presents a complete picture of the sample. Further, the forest plot provides foundational details such that one can observe which analyzed studies exist in each of the frequency distribution bins summarized in Table \ref{table:3}.

\subsection{Phase 2: Subgroup Analysis}
Phase 2 consisted of between-approaches analysis organized by ML model subgroup. Doing so was intended to reveal the potential variability and distribution of accuracy values within the set of misinformation detection techniques. Further, between-approaches analysis was an appropriate mechanism to derive an answer to the second research question. To accomplish this, phase 2 employed the same descriptive analysis as phase 1 albeit iteratively for each ML models subgroup (Table \ref{table:4}). 

\begin{table*}[!htbp]
  \centering
  \begin{threeparttable} \caption{Subgroup summary organized by ML approach}
      \begin{tabular}{cccccccccc}
          \toprule
          \textbf{Approach} & \textbf{Mean}  & \textbf{Lower} & \textbf{Upper} & \textbf{Std} & \textbf{Median} & \textbf{Q1} & \textbf{Q3} & \textbf{Min} & \textbf{Max} \\
                            & \textbf{Accuracy} & \textbf{CI} & \textbf{CI} & \textbf{Dev} & \textbf{Accuracy} & & & \textbf{Accuracy} & \textbf{Accuracy} \\
          \hline
          \hline
          HAN	& 0.95	& 0.939	& 0.960	& 0.014	& 	0.95	& 	0.947	& 	0.956	& 	92.25\%	& 	96.77\% \\
          DNN	& 0.89	& 	0.769	& 	1.009	& 	0.106	& 	0.91	& 	0.842	& 	0.947	& 	77.45\%	& 	98.36\% \\
          GCN	& 0.86	& 0.736	& 	0.981	& 	0.166	& 	0.89	& 	0.883	& 	0.952	& 	49.20\%	& 96.10\% \\
          LSTM & 0.85	& 	0.752	& 	0.941	& 	0.188	& 	0.91	& 	0.815	& 	0.957	& 	23.30\%	& 	99.00\% \\
          ANN	& 0.84	& 	0.824	& 	0.853	& 	0.018	& 	0.84	& 	0.825	& 	0.849	& 	81.60\%	& 	86.50\% \\
          RNN	& 0.82	& 	0.728	& 	0.904	& 	0.149	& 	0.84	& 	0.761	& 	0.934	& 	50.00\%	& 	98.00\% \\
          CNN	& 0.79		& 0.693	& 	0.880	& 	0.197		& 0.86		& 0.723	& 	0.927	& 	27.00\%	& 	99.00\% \\
          BERT	& 0.78	& 	0.612	& 	0.949	& 	0.172	& 	0.77	& 	0.718	& 	0.835	& 	57.90\%	& 99.89\% \\
          DT	& 0.77		& 0.634	& 	0.914	& 	0.160	& 	0.75	& 	0.716	& 	0.890	& 	55.00\%	& 96.00\% \\
          RF	& 0.76	& 	0.598	& 	0.923	& 	0.166	& 	0.80	& 	0.710	& 	0.847	& 53.00\%	& 92.00\% \\
          NB	& 0.74	& 	0.653	& 	0.825	& 	0.164	& 	0.78	& 	0.619	& 	0.874	& 	38.00\%	& 96.08\% \\
          SVM	& 0.71	& 	0.606	& 	0.812	& 	0.204	& 	0.73	& 	0.583	& 	0.877	& 	25.50\%	& 99.90\% \\
          k-NN	& 0.68	& 0.611	& 	0.759	& 	0.092	& 	0.70	& 	0.625	& 	0.706	& 	57.00\%	& 83.00\% \\
          LR	& 0.67	& 0.543	& 0.789	& 0.209	& 0.6	& 0.548	& 0.811	& 25.50\%	& 99.00\% \\
          LLM	& 0.66	& 0.454	& 0.863	& 0.209	& 0.70	& 0.542	& 0.819	& 39.25\%	& 83.60\% \\
          \bottomrule
      \end{tabular}
      \label{table:4}
  \end{threeparttable}
\end{table*}

As an aside, the subgroups were established during data analysis through pattern matching. For example, Hierarchical Attention Networks (HAN) is a proxy for any ML-based disinformation technique in the sample containing the substring “HAN”. Likewise, LSTM proxies for all instances of the substring “LSTM” and so on.

Overall, the subgroup analysis appears to align with the trend indicated in Table \ref{table:3} from Phase 1 insofar as the Mean accuracies across subgroups exist within the frequency distribution majority. Further, the ranges for minimum and maximum accuracies appear to align with min and max values shown in Table \ref{table:1} as well as the distribution shown in Table \ref{table:2}. However, the subgroup summary adds interesting detail.

HAN as a subgroup demonstrated the largest statistical Mean. Opposite, LLM demonstrated the smallest. Additionally, Standard Deviation reveals a low dispersion in the case of HAN techniques and a wide dispersion with LLM. Naturally, high levels of accuracy dispersion appear when techniques have a large gap between minimum and maximum accuracy values. Thus, accuracy alone is but one part of an answer to the second research question. The entirety of the descriptive summary for any given subgroup is what yields the complete picture.

With that stated, one must exercise caution in interpreting the results in comparison to Phase 1. The subgroup summary excluded techniques not belonging to one of the stated subgroups. Such techniques had a frequency of one (e.g., the technique appeared only once in the sample) and thus demonstrated a skewed summary. For those subgroups with high frequency (Figure \ref{fig:1}), one should consider not just Mean but also the CI range in the context of the minimum and maximum reported accuracies. Whereas one specific HAN detection technique might be as relatively strong as any other (the ranges are narrow), SVM for instance might perform exceptionally well in one case (max = 99.9\%) but many more SVM techniques have low detection accuracies (min = 25.50\%, Mean = 71\%).

\subsection{Phase 3: Statistical Tests}
The last phase of the meta-analysis consisted of statistical tests and comparative inference. The phase was augmentative to Phase 2 and did not seek to address a different research question. Instead, this phase illuminated potential within-approaches differences adjacent to between-approaches. With that in mind, an omnibus ANOVA was used to determine if a statistically significant difference existed between the ML-based detection technique subgroups. More specifically, the statistical test sought to determine whether any significant difference exists between reported accuracies in the sample and the associated disinformation detection approaches (e.g. Model types). 

\begin{table*}[!htbp]
  \centering
  \begin{threeparttable} \caption{Results from omnibus ANOVA statistical test}
      \begin{tabular}{lcccc}
          \toprule
          \textbf{Subgroups} & \textbf{Sum of Squares} & \textbf{$df$} & \textbf{$f$} & \textbf{$PR (>F)$} \\
          \hline
          \hline
          HAN, DNN, GCN, LSTM, ANN, RNN, CNN, BERT, & 355.378 & 11.0 & 0.204 & 0.997 \\
          DT, RF, NB, SVM, k-NN, LR, LLM, Other & & & & \\
          Residual & 2366.793 & 141.0 & & \\
          \bottomrule
      \end{tabular}
      \label{table:5}
  \end{threeparttable}
\end{table*}

The ANOVA test revealed the differences in Mean accuracy across the 12 approach categories were not statistically significant ($p$ = 0.997, $\alpha$ = 0.05). Further, the sum of squares for the Approach Subgroups was 355.38, whereas the residual sum of squares was 2,366.79. Such suggested the between‐approach variability (attributable to different ML model types) was minimal compared to the within‐approach variability. The latter directly augments the observations from Phase 2 wherein the results demonstrated high standard deviation within subgroups (e.g. SVM) while many subgroups had Mean accuracies within a tight band. In practical terms, no approach subgroup appeared to outperform any other subgroup in terms of cyber information influence detection effectiveness. However, material differences exist within some subgroups. Based on these outcomes, there was no basis to run a Tukey’s HSD Test. 

\section{Conclusions and Future Work}
Digital communication platforms- social media, discussion forums, and so forth- are besieged by cyber information influence. The effects are wide-ranging with implications to public opinion, institutional trust, and social cohesion [56, 36]. Consequently, researchers have tested various ML techniques to detect information influence, particularly malicious forms. Yet, although the field has demonstrated significant advancements the literature, and results therein, remain fragmented. There is little data to substantiate the overall effectiveness ML-based detection offers. Additionally, inconsistencies in datasets, evaluation methodologies, and feature engineering techniques contribute to the lack of consensus regarding a generalized effective misinformation detection strategy [33, 49].

For these reasons, this study set out to investigate cyber information influence detection effectiveness and explore potential differences between ML-based approaches. A three-phased meta-analysis reviewed 40 papers yielding 153 discrete accuracy metrics stemming from 81 ML models evaluated against 37 datasets. As part of the meta-analysis design, this study posed two research questions. One question was, \textit{what is the overall effectiveness, measured in model accuracy, of ML-based cyber information influence detection}? The first phase of the meta-analysis revealed an answer of 78.19\% accuracy in the sampled literature with a population value between 75\% and 80\%. Meanwhile, the reported maximum of 99.9\% accuracy is notable as it represents a benchmark frontier in the field. The second was, \textit{what is the effectiveness, measured in model accuracy, of each ML approach subgroup and to what extent are there significant differences between approaches}? The second phase of the meta-analysis revealed no significant difference exists between subgroups. Yet, the data suggest within a subgroup there can be great variation in accuracy.

Overall, one can reasonably conclude ML-based cyber information influence detection (a) offers a variety of viable techniques and (b) is particularly effective within the assumptions and limitations of the analyzed literature. At the same time, the data show there cannot be a simple, clear-cut answer to an inquiry into what might be the best detection technique. Throughout the sample literature, one observes tight coupling between a given technique, the training material, and the target application. Recall the sample encompassed 37 different training datasets. Further, discrete detection techniques within a subgroup may significantly vary in accuracy. Thus, it is not possible now to state SVM is better than CNN. A stronger postulate would be to claim a given detection technique tends to exhibit a given accuracy under specific conditions. 

Along these lines, the data did not provide evidence that any approach category (e.g., CNN, LSTM, SVM, etc.) systematically outperforms the others in terms of average accuracy. Indeed, practical factors such as the large within‐category variability, differences in datasets, missing sample sizes, or heterogeneous model implementations might be overshadowing any true ML-based detection technique differences. Moreover, the literature and the results of this study do not address why one might select any given ML-based approach compared to another.

\subsection{Potential Biases and Limitations}
Granted, a limitation of this study is the reliance on existing literature and the veracity of reported findings. It is possible research exists which was not included in the sample that might fill in some indicated gaps. Similarly, the existing literature does not seem to have advanced quite as far as ML or the broader AI field. For example, frontier technologies such as agentic AI and neurosymbolic AI are not yet represented in the research. Whilst neither may end up as appropriate implementations for cyber information influence detection, the lack of presence still represents a limitation of this work. Finally, this work is limited insofar as one must assume there are no errors in the sample whether such be in the implementation of a ML model, the training data for the implementation, as well as in the calculation of accuracy as a performance metric. The limitation is somewhat ameliorated by having restricted the sample to peer reviewed sources. 

Another potential limitation of this meta-analysis is publication bias, wherein studies reporting high detection accuracy are more likely to be published than those with lower or negative findings. This bias may artificially inflate the observed mean accuracy across ML-based cyber information influence detection techniques. To mitigate this effect, future research should consider applying statistical corrections such as Egger’s test or trim-and-fill analysis to estimate the extent of bias and adjust results accordingly. Additionally, while this study systematically selected papers from reputable peer-reviewed sources, the variability in dataset selection across studies remains an inherent challenge. The 40 reviewed studies used 37 different datasets, which introduces significant variation in evaluation conditions. Standardizing dataset selection and evaluation methodologies would significantly enhance comparability in future misinformation detection research.

Furthermore, while this study applied rigorous inclusion criteria, the lack of standardized reporting of feature selection and hyperparameter tuning across ML implementations remains a challenge. Even when studies employed the same ML technique, architectural choices (e.g., activation functions, learning rates, optimization strategies) were not always documented. Future meta-analyses should encourage open-source model implementations and hyperparameter disclosure to improve reproducibility and facilitate deeper insights into performance variations.

\subsection{Recommendations and Future Work}
Principally, additional post-hoc analysis is warranted for existing ML-based cyber information influence detection techniques. On one hand, while this meta-analysis provides a hopefully unifying perspective of the field, there could be value in analyzing specific subgroups more deeply. HAN-based techniques are one such example given the demonstrated accuracy and cohesion around the central tendency measure. As well, where a given detection technique had limited representation in the sample (e.g., DNN, LLM), follow-up work later may be beneficial when those ML-based models reach higher saturation in the literature.

The next recommendation concerns the literature. One can observe the analyzed literature is not equal with respect to technical details. In other words, not all analyzed papers include development and engineering details related to the ML implementation. Few papers offer or provide links to source code. Similarly, even when two or more studies used identical ML models, one cannot be certain of configuration details such as hyperparameters, selected features, or potential fine-tuning. Accordingly, reproduction and replication are possible avenues for future research. Such work would serve to fill in existing gaps in technical details. Moreover, to the extent specific studies can be reproduced or replicated, it would be beneficial to conduct ablative experiments as a way to tease out what elements of the system (i.e., ML algorithm, training data, feature selection, hyperparameters, etc.) affect performance metric results.

As well, the analyzed ML approaches tested on different datasets. Doing so potentially masks differences in raw accuracy. Thus, it would be beneficial to measure accuracy (and additional performance metrics) across subgroups whilst maintaining a single dataset. There also may be benefit in testing a single ML model subgroup across a variety of datasets. The former may establish a more robust effectiveness for general ML-based misinformation detection whereas the latter quantifies the level of generalizability associated with a given ML model type.

A critical point to consider is existing disinformation detection techniques rely on generated information. Meaning, whether the disinformation is human or AI generated, the detection schemes operate on generated artifacts (e.g., posts or tweets). Thus, one issue is the artifact is already in the wild before detection can be applied. In some sense, the damage to the targeted information ecosystem has been achieved by that point. Another issue is the existing detection methods do not tell us anything about what is contained within a model. As much as the contents of a model are controlled in the sample studies because of the controlled nature of the research, access to and use of publicly accessible models is on the rise. Future work, then, might be fitting if misinformation detection methods can peer into such models.

\section{Final Remarks}
This study presents the first large-scale meta-analysis quantifying the effectiveness of ML-based cyber information influence detection techniques. The findings provide valuable insights for researchers, policymakers, and practitioners, emphasizing that no single model is universally superior and that methodological factors significantly impact detection performance. To advance the field, future research should focus on standardization, frontier technique innovation, open-access implementations, adversarial robustness, and interdisciplinary collaboration. By addressing the existing challenges, the cyber information influence detection research community can contribute to the development of more transparent, effective, and ethically responsible techniques that better serve the public interest.

\section{Funding}
This work was supported by the Australian Fulbright Commission and the University of Adelaide. The work began and was completed in conjunction with a Fulbright Scholar award whilst in residence at the University of Adelaide.

%\bibliographystyle{apacite}
%\bibliography{references.bib}
\section*{References}
\begin{hangparas}{.25in}{1}
  [1] Adhvaryu, K. U. (2024). A Survey of Fake Data or Misinformation Detection Techniques Using Big Data and Sentiment Analysis. SN Computer Science, 5(7), 1-16.

  [2] Aghakhani, H., Machiry, A., Nilizadeh, S., Kruegel, C., \& Vigna, G. (2018, May). Detecting deceptive reviews using generative adversarial networks. In 2018 IEEE security and privacy workshops (SPW) (pp. 89-95). IEEE.
  
  [3] Agrawal, T., Gupta, R., \& Narayanan, S. (2017, August). Multimodal detection of fake social media use through a fusion of classification and pairwise ranking systems. In 2017 25th European Signal Processing Conference (EUSIPCO) (pp. 1045-1049). IEEE.
  
  [4] Agarwal, A., \& Dixit, A. (2020, May). Fake news detection: an ensemble learning approach. In 2020 4th International Conference on Intelligent Computing and Control Systems (ICICCS) (pp. 1178-1183). IEEE.
  
  [5]  Ahmed, H., Traore, I., \& Saad, S. (2017). Detection of online fake news using n-gram analysis and machine learning techniques. In Intelligent, Secure, and Dependable Systems in Distributed and Cloud Environments: First International Conference, ISDDC 2017, Vancouver, BC, Canada, October 26-28, 2017, Proceedings 1 (pp. 127-138). Springer International Publishing.
  
  [6] Ajao, O., Bhowmik, D., \& Zargari, S. (2018, July). Fake news identification on twitter with hybrid cnn and rnn models. In Proceedings of the 9th international conference on social media and society (pp. 226-230).
  
  [7] Al-Rawi, A., \& Fakida, A. (2023). The methodological challenges of studying “fake news”. Journalism Practice, 17(6), 1178-1197.
  
  [8] Aphiwongsophon, S., \& Chongstitvatana, P. (2018, July). Detecting fake news with machine learning method. In 2018 15th international conference on electrical engineering/electronics, computer, telecommunications and information technology (ECTI-CON) (pp. 528-531). IEEE.
  
  [9] Bahad, P., Saxena, P., \& Kamal, R. (2019). Fake news detection using bi-directional LSTM-recurrent neural network. Procedia  Computer Science, 165, 74-82.
  
  [10] Bian, T., Xiao, X., Xu, T., Zhao, P., Huang, W., Rong, Y., \& Huang, J. (2020, April). Rumor detection on social media with bi-directional graph convolutional networks. In Proceedings of the AAAI conference on artificial intelligence (Vol. 34, No. 01, pp. 549-556).
  
  [11] Chen, C., \& Shu, K. (2024). Combating misinformation in the age of llms: Opportunities and challenges. AI Magazine, 45(3), 354-368.
  
  [12] Choudhary, M., Jha, S., Saxena, D., \& Singh, A. K. (2021, May). A review of fake news detection methods using machine learning. In 2021 2nd international conference for emerging technology (INCET) (pp. 1-5). IEEE.
  
  [13] Elmurngi, E., \& Gherbi, A. (2017). Detecting fake reviews through sentiment analysis using machine learning techniques. IARIA/data analytics, 65-72.
  
  [14] Fang, Y., Wang, H., Zhao, L., Yu, F., \& Wang, C. (2020). Dynamic knowledge graph based fake-review detection. Applied Intelligence, 50, 4281-4295.
  
  [15] Ghosal, D., Bhatnagar, S., Akhtar, M. S., Ekbal, A., \& Bhattacharyya, P. (2017, August). IITP at SemEval-2017 task 5: an ensemble of deep learning and feature based models for financial sentiment analysis. In Proceedings of the 11th international workshop on semantic evaluation (SemEval-2017) (pp. 899-903).
  
  [16] Granik, M., \& Mesyura, V. (2017, May). Fake news detection using naive Bayes classifier. In 2017 IEEE first Ukraine conference on electrical and computer engineering (UKRCON) (pp. 900-903). IEEE.
  
  [17] Hardalov, M., Arora, A., Nakov, P., \& Augenstein, I. (2022). A survey on stance detection for mis-and disinformation identification. In Findings of the Association for Computational Linguistics: NAACL 2022, 1259-1277.
  
  [18] Hasan, A., Moin, S., Karim, A., \& Shamshirband, S. (2018). Machine learning-based sentiment analysis for twitter accounts. Mathematical and computational applications, 23(1), 11.
  
  [19] Hiramath, C. K., \& Deshpande, G. C. (2019, July). Fake news detection using deep learning techniques. In 2019 1st International Conference on Advances in Information Technology (ICAIT) (pp. 411-415). IEEE.
  
  [20] Howells, K., \& Ertugan, A. (2017). Applying fuzzy logic for sentiment analysis of social media network data in marketing. Procedia computer science, 120, 664-670.
  
  [21] Hu, G., Ding, Y., Qi, S., Wang, X., \& Liao, Q. (2019). Multi-depth graph convolutional networks for fake news detection. In Natural Language Processing and Chinese Computing: 8th CCF International Conference, NLPCC 2019, Dunhuang, China, October 9–14, 2019, Proceedings, Part I 8 (pp. 698-710). Springer International Publishing.
  
  [22] Huang, G., Jia, W., \& Yu, W. (2024). Media Literacy Interventions Improve Resilience to Misinformation: A Meta-Analytic Investigation of Overall Effect and Moderating Factors. Communication Research, 00936502241288103.
  
  [23] Ibrishimova, M. D., \& Li, K. F. (2020). A machine learning approach to fake news detection using knowledge verification and natural language processing. In Advances in Intelligent Networking and Collaborative Systems: The 11th International Conference on Intelligent Networking and Collaborative Systems (INCoS-2019) (pp. 223-234). Springer International Publishing.
  
  [24] Islam, N., Shaikh, A., Qaiser, A., Asiri, Y., Almakdi, S., Sulaiman, A., Moazzam, V., \& Babar, S. A. (2021). Ternion: An autonomous model for fake news detection. Applied Sciences, 11(19), 9292.
  
  [25] Jadhav, S. S., \& Thepade, S. D. (2019). Fake news identification and classification using DSSM and improved recurrent neural network classifier. Applied Artificial Intelligence, 33(12), 1058-1068.
  
  [26] Kaliyar, R. K., Goswami, A., Narang, P., \& Sinha, S. (2020). FNDNet–a deep convolutional neural network for fake news detection. Cognitive Systems Research, 61, 32-44.
  
  [27] Kang, Z., Cao, Y., Shang, Y., Liang, T., Tang, H., \& Tong, L. (2021, May). Fake news detection with heterogenous deep graph convolutional network. In Pacific-Asia Conference on Knowledge Discovery and Data Mining (pp. 408-420). Cham: Springer International Publishing.
  
  [28] Kansara, P., \& Adhvaryu, K. U. (2024). A Survey of Fake Data or Misinformation Detection Techniques Using Big Data and Sentiment Analysis. SN Comput. Sci., 5(7). doi:10.1007/s42979-024-03297-z
  
  [29] Kareem, W., \& Abbas, N. (2023, November). Fighting lies with intelligence: Using large language models and chain of thoughts technique to combat fake news. In International Conference on Innovative Techniques and Applications of Artificial Intelligence (pp. 253-258). Cham: Springer Nature Switzerland.
  
  [30] Kaur, S., Kumar, P., \& Kumaraguru, P. (2020). Automating fake news detection system using multi-level voting model. Soft Computing, 24(12), 9049-9069.
  
  [31] Kula, S., Choraś, M., \& Kozik, R. (2021). Application of the bert-based architecture in fake news detection. In 13th International Conference on Computational Intelligence in Security for Information Systems (CISIS 2020) 12 (pp. 239-249). Springer International Publishing.
  
  [32] Kumar, R., Goddu, B., Saha, S., \& Jatowt, A. (2024). Silver lining in the fake news cloud: Can large language models help detect misinformation?. IEEE Transactions on Artificial Intelligence, 6(1), 14-24.
  
  [33] Kumar, S., Asthana, R., Upadhyay, S., Upreti, N., \& Akbar, M. (2020). Fake news detection using deep learning models: A novel approach. Transactions on Emerging Telecommunications Technologies, 31(2), e3767.
  
  [34] Mahabub, A. (2020). A robust technique of fake news detection using Ensemble Voting Classifier and comparison with other classifiers. SN Applied Sciences, 2(4), 525.
  
  [35] Mahmud, Y., Shaeeali, N. S., \& Mutalib, S. (2021). Comparison of machine learning algorithms for sentiment classification on fake news detection. International Journal of Advanced Computer Science and Applications, 12(10).
  
  [36] Mashru, D., \& Nautiyal, N. S. (2024). Unveiling the Truth: A Literature Review on Leveraging Computational Linguistics for Enhanced Forensic Analysis. In International Semantic Intelligence Conference (pp. 71-84). Springer, Singapore.
  
  [37] Menczer, F., Crandall, D., Ahn, Y. Y., \& Kapadia, A. (2023). Addressing the harms of AI-generated inauthentic content. Nature Machine Intelligence, 5(7), 679-680.
  
  [38] Mridha, M. F., Keya, A. J., Hamid, M. A., Monowar, M. M., \& Rahman, M. S. (2021). A comprehensive review on fake news detection with deep learning. IEEE access, 9, 156151-156170.
  
  [39] Nasir, J. A., Khan, O. S., \& Varlamis, I. (2021). Fake news detection: A hybrid CNN-RNN based deep learning approach. International Journal of Information Management Data Insights, 1(1), 100007.
  
  [40] Oshikawa, R., Qian, J., \& Wang, W. Y. (2020, May). A Survey on Natural Language Processing for Fake News Detection. In Proceedings of the Twelfth Language Resources and Evaluation Conference (pp. 6086-6093).
  
  [41] Papageorgiou, E., Chronis, C., Varlamis, I., \& Himeur, Y. (2024). A survey on the use of large language models (llms) in fake news. Future Internet, 16(8), 298.
  
  [42] Puska, A. A., Baroni, L. A., \& Pereira, R. (2024). Decoding the Sociotechnical Dimensions of Digital Misinformation: A Comprehensive Literature Review. arXiv preprint arXiv:2406.11853.
  
  [43] Qian, S., Hu, J., Fang, Q., \& Xu, C. (2021). Knowledge-aware multi-modal adaptive graph convolutional networks for fake news detection. ACM Transactions on Multimedia Computing, Communications, and Applications (TOMM), 17(3), 1-23.
  
  [44] Ramadhani, A. M., \& Goo, H. S. (2017, August). Twitter sentiment analysis using deep learning methods. In 2017 7th International annual engineering seminar (InAES) (pp. 1-4). IEEE.
  
  [45] Ren, Y., \& Zhang, Y. (2016, December). Deceptive opinion spam detection using neural network. In Proceedings of COLING 2016, the 26th International Conference on Computational Linguistics: Technical Papers (pp. 140-150).
  
  [46] Roy, P. K., \& Chahar, S. (2020). Fake profile detection on social networking websites: a comprehensive review. IEEE Transactions on Artificial Intelligence, 1(3), 271-285.
  
  [47] Sharma, D. K., \& Garg, S. (2023). IFND: a benchmark dataset for fake news detection. Complex \& intelligent systems, 9(3), 2843-2863.
  
  [48] Shu, K., Sliva, A., Wang, S., Tang, J., \& Liu, H. (2017). Fake news detection on social media: A data mining perspective. ACM SIGKDD explorations newsletter, 19(1), 22-36.
  
  [49] Shu, K., Mahudeswaran, D., Wang, S., Lee, D., \& Liu, H. (2020). Fakenewsnet: A data repository with news content, social context, and spatiotemporal information for studying fake news on social media. Big data, 8(3), 171-188.
  
  [50] Singhania, S., Fernandez, N., \& Rao, S. (2017). 3han: A deep neural network for fake news detection. In Neural Information Processing: 24th International Conference, ICONIP 2017, Guangzhou, China, November 14-18, 2017, Proceedings, Part II 24 (pp. 572-581). Springer International Publishing.
  
  [51] Su, J., Cardie, C., \& Nakov, P. (2024, June). Adapting Fake News Detection to the Era of Large Language Models. In Findings of the Association for Computational Linguistics: NAACL 2024 (pp. 1473-1490).
  
  [52] Sun, Y., He, J., Cui, L., Lei, S., \& Lu, C. T. (2024). Exploring the Deceptive Power of LLM-Generated Fake News: A Study of Real-World Detection Challenges. arXiv preprint arXiv:2403.18249.
  
  [53] Tacchini, E., Ballarin, G., Della Vedova, M. L., Moret, S., \& De Alfaro, L. (2017). Some like it hoax: Automated fake news detection in social networks. arXiv preprint arXiv:1704.07506.
  
  [54] Tajrian, M., Rahman, A., Kabir, M. A., \& Islam, M. R. (2023). A review of methodologies for fake news analysis. IEEE Access.
  
  [55] Tanvir, A. A., Mahir, E. M., Akhter, S., \& Huq, M. R. (2019, June). Detecting fake news using machine learning and deep learning algorithms. In 2019 7th international conference on smart computing \& communications (ICSCC) (pp. 1-5). IEEE.
  
  [56] Varlamis, I., Michail, D., Glykou, F., \& Tsantilas, P. (2022). A survey on the use of graph convolutional networks for combating fake news. Future Internet, 14(3), 70.
  
  [57] Wang, W. Y. (2017). " Liar, Liar Pants on Fire": A New Benchmark Dataset for Fake News Detection. In Proceedings of the 55th Annual Meeting of the Association for Computational Linguistics (Volume 2: Short Papers). Association for Computational Linguistics.
  
  [58] Wang, L., Wang, Y., De Melo, G., \& Weikum, G. (2019). Understanding archetypes of fake news via fine-grained classification. Social Network Analysis and Mining, 9, 1-17.
  
  [59] Wang, Y., Qian, S., Hu, J., Fang, Q., \& Xu, C. (2020, June). Fake news detection via knowledge-driven multimodal graph convolutional networks. In Proceedings of the 2020 international conference on multimedia retrieval (pp. 540-547).
  
  [60] Wu, J., Guo, J., \& Hooi, B. (2024, August). Fake News in Sheep's Clothing: Robust Fake News Detection Against LLM-Empowered Style Attacks. In Proceedings of the 30th ACM SIGKDD conference on knowledge discovery and data mining (pp. 3367-3378).
  
  [61] Xu, D., Fan, S., \& Kankanhalli, M. (2023, October). Combating misinformation in the era of generative AI models. In Proceedings of the 31st ACM International Conference on Multimedia (pp. 9291-9298).
  
  [62] Zhang, X., \& Ghorbani, A. A. (2020). An overview of online fake news: Characterization, detection, and discussion. Information Processing \& Management, 57(2), 102025.
  
  [63] Zhou, J., Zhang, Y., Luo, Q., Parker, A. G., \& De Choudhury, M. (2023, April). Synthetic lies: Understanding ai-generated misinformation and evaluating algorithmic and human solutions. In Proceedings of the 2023 CHI Conference on Human Factors in Computing Systems (pp. 1-20).
  
  [64] Zhou, X., Zafarani, R., Shu, K., \& Liu, H. (2019, January). Fake news: Fundamental theories, detection strategies and challenges. In Proceedings of the twelfth ACM international conference on web search and data mining (pp. 836-837).
  
  [65] Zubiaga, A., Aker, A., Bontcheva, K., Liakata, M., \& Procter, R. (2018). Detection and resolution of rumours in social media: A survey. Acm Computing Surveys (Csur), 51(2), 1-36.

\end{hangparas}

%% appendix A
\pagebreak
\section*{\centering{Appendix A}}
\label{appendixA}

\begin{table*}[!h]
  \small
  \centering
  \begin{threeparttable} %\caption{Details of literature analyzed during the meta-analysis}
      \begin{longtable}{c|c|p{3.0in}|p{1.5in}}
          \toprule
          \textbf{Review} & \textbf{Study}  & \textbf{ML Models}  & \textbf{Datasets} \\
          \hline
          \hline
          1 & 2 & GANs & Ott et al. (2011)$^1$ \\
          1 & 3 &  Pairwise Ranking & \\
          1 & 23 & k-NN & Kaggle Fake News \\
          1 & 13 &  NB, SVM, k-NN, K*, DT & Movie reviews 1.0 and 2.0 \\
          1 & 14 & CRF, BRNN, LSTM, ST-BLSTM & Cui (2019)$^2$ \\
          1 & 15 & CNN-LSTM & Twitter \\
          1 & 16 & NB & Buzzfeed News \\
          1 & 20 & Fuzzy Logic & Twitter \\
          1 & 25 & DSSM-LSTM & LIAR \\
          1 & 26 & DNN & Kaggle News \\
          1 & 30 & Linear SVC & News Trends, Kaggle Fake News, Reuters \\
          1 & 34 & Ensemble Voting Classifier & FakeNewsNet \\
          1 & 44 & DNN & Twitter, Facebook \\         

          12 & 4 & SVM, NB, k-NN, CNN, LSTM & LIAR + Kaggle \\
          12 & 5 & SVM, LSVM, k-NN, DT, SGD, LR & 2016 US Election articles \\
          12 & 8 & NB, NN, SVM & Twitter \\
          12 & 57 & SVM, LR, LSTM, NB, RNN & Twitter Chile Earthquake \\
          
          28 & 6 & LSTM, LSTMDrop, LSTM-CNN & Zubiaga et al. (65) \\
          28 & 45 & CNN, RNN, GRNN, Bidirectional GRNN, Bidirectional GRRN & Li et al. (2014)$^3$ \\ 
          28 & 62 & Hybrid deep learning models & SHPT-4 class, Politi-6 class, Politi-4 class \\

          35 & 18 & NB TextBlob, NB Sentiwordnet, NB WSD, SVM TextBlob, SVM Sentiwordnet, SVM WSD & Twitter \\
          35 & 19 & LR, RF, SVM, NB, DNN & News Articles \\

          38 & 31 & BERT & ISOT \\
          38 & 33 & CNN, LSTM, Bidirectional LSTM, CNN + LSTM ensemble, Bidirectional LSTM + LSTM ensemble, CNN + LSTM with attention, CNN + Bidirectional LSTM with attention, LR, SVM & FakeNewsNet \\
          38 & 53 & GloVe-Ave, GRU, GRU-Ave, HAN-Ave, HAN-Max, HAN, 3HAN-Ave, 3HAN-Max, 3HAN, 3HAN+PT & PolitiFact, Forbes \\
          38 & 56 & EANN, MVAE, BDANN-v, BDANN & Weibo, Weibo Filtered \\

          40 & 52 & SVM, LR, NB, CNN & PolitiFact, GossipCop \\
          40 & 55 & LR, Harmonic BLC & Facebook \\
          40 & 59 & SVM, LR, Bi-LSTM, CNN & LIAR \\

          41 & 9 & CNN, Vanilla RNN, Unidirectional LSTM-RNN, Bidirectional LSTM-RNN & DS1, DS2 \\
          41 & 29 & Fine-tuned Llama2 & LIAR \\
          41 & 30 & LR, RF, MNB, SGD, KNNs, DT, AB, CNN, RNN, Hybrid CNN-RNN & FA-KES \\
          41 & 54 & Fine-tuned PLM (BERT, RoBERTa, FnBERT) Trained SOTA (Grover, DualEmo), Fine-tuned LLM Llama2-7b+LoRA, Fine-tuned LLM Vicuna-7b+LoRA, Comm. LLM ChatGPT-3.5 & VLFPN \\

          49 & 24 & DT, RF, LR, SVM & Kaggle \\

          58 & 10 & Bi-directional GCN & Weibo, Twitter15, Twitter16 \\
          58 & 21 & Multi-depth GCN & LIAR \\
          58 & 27 & Heterogeneous GCN & Fakeddit \\
          58 & 43 & GCN + VGG-19 for visual processing & Weibo \\
          58 & 55 & Knowledge-driven multimodal GCN & Weibo \\

          \bottomrule
      \end{longtable}
      %\label{table:5}
      \begin{tablenotes}
        \item[1] M. Ott, Y. Choi, C. Cardie, and J. T. Hancock, “Finding deceptive opinion spam by any stretch of the imagination,” in Proceedings of the 49th Annual Meeting of the Association for Computational Linguistics: Human Language Technologies-Volume 1. Association for Computational Linguistics, 2011, pp. 309–319.
        \item[2] Cui, J. (2019). https://github.com/Butterfler/fake-review/tree/master
        \item[3] Li, J., Ott, M., Cardie, C., \& Hovy, E. (2014, June). Towards a general rule for identifying deceptive opinion spam. In Proceedings of the 52nd Annual Meeting of the Association for Computational Linguistics (Volume 1: Long Papers) (pp. 1566-1576).
      \end{tablenotes}
  \end{threeparttable}
\end{table*}

%% appendix B
\pagebreak
\section*{\centering{Appendix B}}
\label{appendixB}

\begin{figure}[!h]
  %\centering
  %\caption{The distribution of ML models in the studies analyzed}
  \label{appendix:2}
  \includegraphics[scale=0.5]{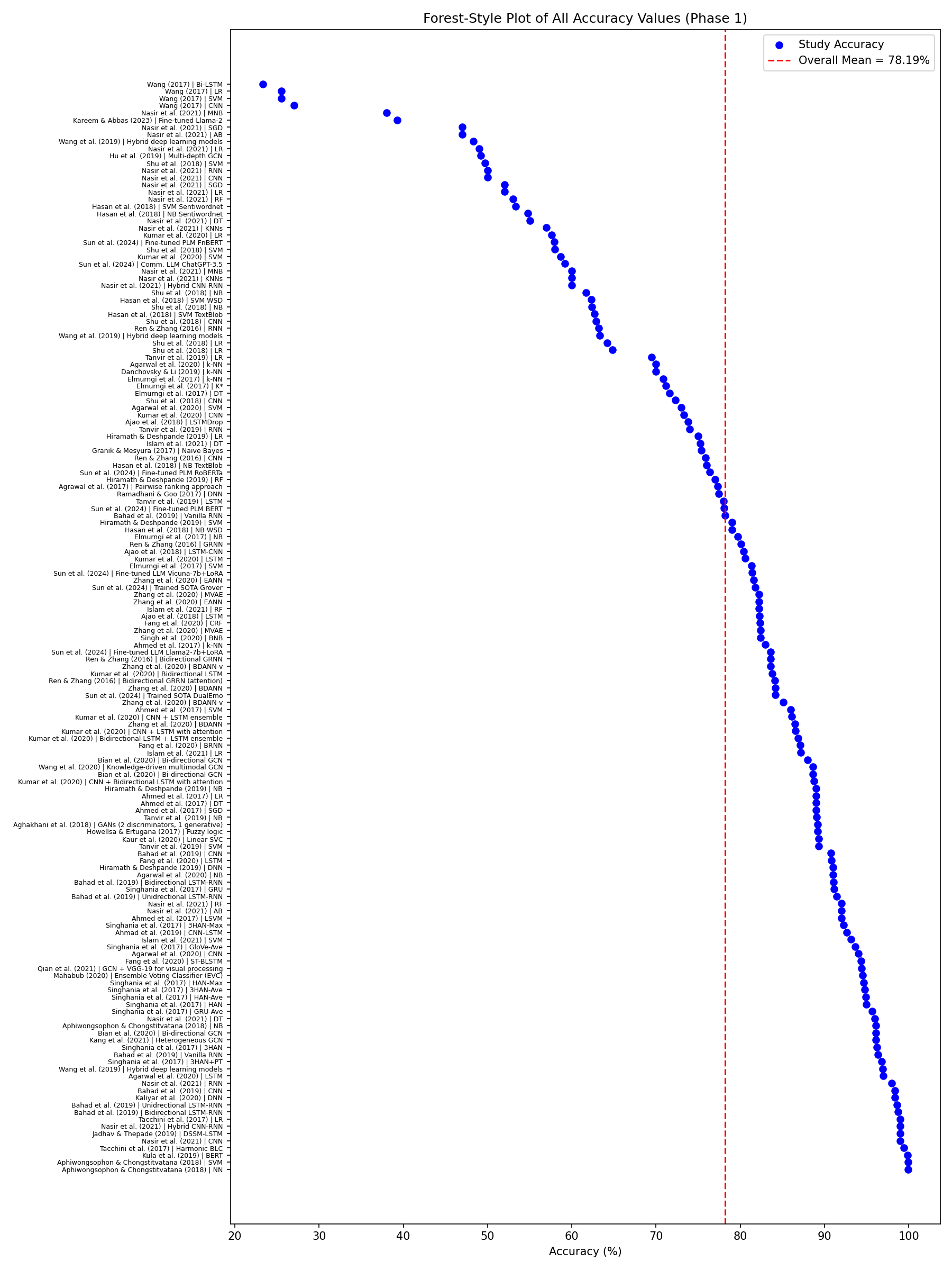}
\end{figure}

\end{document}